\documentclass[10pt]{cai26}
\usepackage{wrapfig}
\usepackage{amsmath,amssymb,amsfonts}
\usepackage{algorithmic}
\usepackage{graphicx}
\usepackage{textcomp}
\usepackage{xcolor}
\usepackage[export]{adjustbox}
\usepackage{wrapfig}
\usepackage{multirow}
\usepackage{graphicx}
\usepackage{subcaption}
\usepackage{booktabs}
\usepackage{url}
\usepackage[compatibility=false]{caption}
\usepackage{booktabs,multirow,siunitx}
\def\BibTeX{{\rm B\kern-.05em{\sc i\kern-.025em b}\kern-.08em
    T\kern-.1667em\lower.7ex\hbox{E}\kern-.125emX}}

\begin{document}
% Editorial staff will replace the following values:
% 1. Conference Year
% 2. Issue number
% 3. Article DOI
\def\conferenceyear{2026}
\volumeheader{39}{0}%{00.000}
\begin{center}

\title{Fairness Audits of Institutional Risk Models in Deployed ML Pipelines}
\maketitle

\thispagestyle{empty}
\pagenumbering{gobble}

\begin{tabular}{c}
Kelly M\textsuperscript{c}Convey\upstairs{\affilone,*}, Dipto Das\upstairs{\affilone}, Maya Ghai\upstairs{\affilone}, Angelina Zhai\upstairs{\affilone}, Rosa Lee\upstairs{\affilone}, Shion Guha\upstairs{\affilone}
\\[0.25ex]
{\small \upstairs{\affilone} University of Toronto, Toronto, Ontario, Canada} \\
\end{tabular}
  
\emails{
  \upstairs{*}kelly.mcconvey@mail.utoronto.ca 
}
\vspace*{0.2in}
\end{center}
\begin{abstract}
Fairness audits of institutional risk models are critical for understanding how deployed machine learning pipelines allocate resources. Drawing on multi-year collaboration with Centennial College, where our prior ethnographic work introduced the ASP-HEI Cycle, we present a replica-based audit of a deployed Early Warning System (EWS), replicating its model using institutional training data and design specifications. We evaluate disparities by gender, age, and residency status across the full pipeline (training data, model predictions, and post-processing) using standard fairness metrics. Our audit reveals systematic misallocation: younger, male, and international students are disproportionately flagged for support, even when many ultimately succeed, while older and female students with comparable dropout risk are under-identified. Post-processing amplifies these disparities by collapsing heterogeneous probabilities into percentile-based risk tiers. This work provides a replicable methodology for auditing institutional ML systems and shows how disparities emerge and compound across stages, highlighting the importance of evaluating construct validity alongside statistical fairness. It contributes one empirical thread to a broader program investigating algorithms, student data, and power in higher education.
\end{abstract}
 
% add your keywords
\copyrightnotice
 
%{\noindent{\bf Editors:} Lydia Bouzar-Benlabiod, Carson K. Leung}
\section{Introduction}
 
Public higher education institutions (HEIs) under sustained fiscal pressure have increasingly turned to algorithmic decision-making as a tool for triaging students \cite{mcconvey_risk_2026}. Our prior ethnographic work at Centennial College, a public college in Ontario, Canada, theorized this turn as the \emph{ASP-HEI Cycle}: a self-reinforcing pattern in which the prioritization of financial sustainability drives the adoption of data-driven practices that reproduce and expand institutional power \cite{mcconvey_this_2024}. That work's argument about the \textit{formalization of inequity} (§5.4.2) held that deployed risk models encode existing demographic inequities into institutional policy, because biases in training data, opaque categorization, and the quantitative framing of risk interact with pre-existing institutional inequities to disproportionately harm students from equity-seeking groups.
 
That argument rested on interviews and stakeholder perception; the pipeline-level mechanisms were not, and could not be, tested from ethnographic data alone. At which stages of a deployed Early Warning System (EWS) does inequity enter, and where does it compound? Does the shift from continuous predictions to percentile tiers dampen or magnify demographic disparity? Answering these questions requires something rare in the public-sector ML literature: access to a deployed institutional model together with the training data and design decisions used to build it \cite{raji_closing_2020}.
 
This paper tests the §5.4.2 claim and extends the framework. Through continued collaboration with the college at the center of the original ASP-HEI ethnography, we replicate its deployed EWS using the institution's training data and documented design specifications, then audit the full pipeline for disparities by gender, age, and residency status. Our audit substantiates the claim: each stage of the pipeline is a distinct site of inequity \emph{production} rather than mere reflection, with baseline group differences learned by the model and compounded downstream. Beyond confirming the claim, we identify the specific mechanism of amplification: \emph{percentile-based post-processing} is where baseline demographic differences are converted into amplified institutional policy. We ask:
 
\begin{itemize}
    \item \textbf{RQ1:} How are different population groups represented and treated by the EWS's (a) training data, (b) predictions, and (c) interpretation?
    \item \textbf{RQ2:} How does the EWS operationalize risk in terms of student drop-out, and how does that operationalization mediate the ASP-HEI cycle's inequity claim?
\end{itemize}
 
Our contributions are: (1) \textbf{empirical substantiation of ASP-HEI §5.4.2}, showing with real institutional data that the deployed model formalizes historical disparity as current allocation, and specifying the pipeline stages at which this occurs; (2) an \textbf{extension of the framework} that identifies percentile-based post-processing as a specific, locatable mechanism of amplification; and (3) a \textbf{replicable audit methodology} that generalizes beyond this site to other institutional risk models where training data and design specifications are accessible.
 
\section{Related Work}
 
Research on algorithmic systems in child welfare \cite{saxena_framework_2021} and homelessness \cite{moon_human-centered_2024} has documented how predictive systems perpetuate disparities through problem formulation and target-variable choice. Raji et al. \cite{raji_closing_2020} argue that internal algorithmic audits are essential for accountability, and Simbeck \cite{simbeck_they_2024} proposes concrete criteria---fairness, transparency, and robustness---for auditing learning analytics systems specifically; our work operationalizes such criteria on a deployed institutional EWS. Obermeyer et al. \cite{obermeyer_dissecting_2019} showed how poor proxy choices produce differential mismeasurement across groups; Passi and Barocas \cite{passi_problem_2019} show how problem-formulation decisions embed values; and Jacobs and Wallach \cite{jacobs_measurement_2021} highlight how institutional convenience often takes precedence over construct validity. Within higher education, Perdomo et al. \cite{perdomo_difficult_2023} evaluate Wisconsin's EWS, demonstrating how prediction-based approaches may fail their stated goals due to problem-formulation issues. Our prior systematic review of algorithms in higher education documented a trend toward less interpretable models built on protected attributes and identified a critical lack of empirical study of how these systems function in practice \cite{mcconvey_human-centered_2023}; our own prior ethnographic work at the present site introduced the ASP-HEI Cycle and identified the formalization of inequity as a central mechanism, but rested on interview data \cite{mcconvey_this_2024}. Our audit addresses that gap at the pipeline level. 
% while our prior ethnographic work at the present site introduced the \emph{ASP-HEI Cycle}, in which financial-sustainability pressure drives HEIs to adopt data-driven practices that reproduce institutional power through increased surveillance, exacerbation of existing inequities (\S5.4.2), and automation of the faculty--student relationship \cite{mcconvey_this_2024}. That work identified the formalization of inequity as a central mechanism but rested on interview data; the pipeline-level mechanisms were left open. Our audit addresses that gap by tracing disparity production across the actual pipeline at the original research site.
 
\section{Methods}
% Our research site is Centennial College, the same institution studied in the original ASP-HEI ethnography \cite{mcconvey_this_2024}; the EWS we audit is the same Early Alert System examined ethnographically in that work. Through continued institutional collaboration we obtained 
Through continued collaboration with Centennial College — the same site and same Early Alert System examined in the original ASP-HEI ethnography \cite{mcconvey_this_2024} — we obtained training data and documented design specifications, enabling a \emph{replica-based audit}---a fairness evaluation conducted on a faithful re-implementation of the deployed model, trained on the same institutional data and following the same documented design specifications---with access to actual training data, documented institutional decisions, and the operational context in which fairness is a resource-allocation question affecting real students.
 
\subsection{Data and Modeling} \label{sec:data-modeling}
After research ethics approval, we obtained student records spanning 2011--2019 (102,353 records: 61,375 domestic, 40,978 international). To replicate the intake EWS, we filtered to first-semester students with features available at intake and trained separate XGBoost models for each population, following institutional design; feature sets differ (English test scores for international, high-school grades for domestic) due to distinct admissions processes. We retained 45 intake/program/admissions features after removing those with substantial missing data. Sensitive attributes (age, gender, residency, funding, first-generation status) were included as model features in line with the replicated system; race/ethnicity and disability data were unavailable. Students were labeled \textit{Successful} if they completed their program within the allowable period and \textit{Unsuccessful} otherwise (withdrawal, failure, transfer). Data were split chronologically (70/15/15), balanced using SMOTE \cite{kabir_balancing_2024}, and tuned via grid search.
 
The EWS pipeline has three stages: (1) intake data collection, (2) separate XGBoost models producing success probabilities, and (3) percentile-based categorization into Low Risk (top 50\%), Medium Risk (next 27\%), and High Risk (bottom 23\%). Test predictions yielded 15{,}461 students (9{,}209 domestic, 6{,}252 international) for fairness analysis.
 
\subsection{Fairness Metrics and Risk Operationalization}
We evaluate disparities using Statistical Parity Difference (SPD; difference in positive prediction rates between groups), Equal Opportunity Difference (EOD; difference in true positive rates), False Positive Rate gaps ($\Delta$FPR; difference in false positive rates), and Calibration Error (CE; deviation between predicted probabilities and observed outcome frequencies), computing pairwise disparities and reporting maximum absolute differences \cite{raji_closing_2020}. A central methodological decision is the target definition: the institution's EWS predicts whether a student will be \textit{unsuccessful}, collapsing dropout, transfer, and program change into a single label, treating dropout as a proxy for \textit{student need for support}. Following Jacobs and Wallach \cite{jacobs_measurement_2021}, we treat this as a construct-validity issue, and for this audit we adopt the institution's definition to faithfully replicate the deployed system. We return to its implications in Section \ref{discussion}.
 
\section{Findings}
 
We audit the EWS across three demographic attributes: gender, age, and residency status.
 
\textbf{Baseline and model-level disparities.} Training data reveal significant group-level differences in program completion: international students achieve 85\% success compared with 67\% for domestic ($\chi^2 = 2847.3$, $p<0.001$); female students outperform male in both populations (domestic 73\% vs.\ 59\%; international 89\% vs.\ 82\%, $p<0.001$); and students aged 26+ outperform those $\leq$20. International-student models achieve significantly higher accuracy (91\%) than domestic (82\%, $p<0.001$), translating into differential access to interventions; both models rely heavily on program-level features (credential type, program length), raising questions about whether student-level prediction is necessary when program-level patterns dominate. Table \ref{tab:combined_performance} summarizes error rates across gender and age groups. Female students face systematically higher false-positive rates (domestic: 32\% vs.\ 23\%; international: 26\% vs.\ 18\%) but lower false-negative rates; all metrics exceed commonly used fairness thresholds (SPD $> 0.1$). Age effects are larger: students aged 36--40 show false-positive rates above 0.60 in both models, while students aged 19--20 exhibit false-negative rates of 0.21 in the domestic model.
 
\begin{table*}[ht]
    \centering
    \caption{Model performance by gender and age. FPR = false positive rate; FNR = false negative rate. Accuracy (gender): Dom.\ F 0.84, M 0.81; Intl.\ F 0.92, M 0.91.}
    \label{tab:combined_performance}
    \begin{tabular}{llcccccc}
        \toprule
         & & \multicolumn{3}{c}{\textbf{Domestic}} & \multicolumn{3}{c}{\textbf{International}} \\
         \cmidrule(lr){3-5} \cmidrule(lr){6-8}
         & \textbf{Group} & \textbf{FPR} & \textbf{FNR} & \textbf{F1} & \textbf{FPR} & \textbf{FNR} & \textbf{F1} \\
        \midrule
        \multirow{2}{*}{\textit{Gender}}
         & Female & 0.32 & 0.10 & 0.89 & 0.26 & 0.05 & 0.96 \\
         & Male   & 0.23 & 0.17 & 0.83 & 0.18 & 0.08 & 0.94 \\
        \midrule
        \multirow{7}{*}{\textit{Age}}
         & 0--18  & 0.21 & 0.18 & 0.84 & 0.19 & \textbf{0.09} & 0.93 \\
         & 19--20 & 0.20 & \textbf{0.21} & 0.81 & 0.18 & \textbf{0.10} & 0.92 \\
         & 21--25 & 0.29 & 0.10 & 0.88 & 0.18 & 0.06 & 0.96 \\
         & 26--30 & 0.33 & 0.09 & 0.90 & 0.25 & 0.03 & 0.97 \\
         & 31--35 & 0.41 & 0.07 & 0.92 & 0.29 & 0.05 & 0.96 \\
         & 36--40 & \textbf{0.61} & 0.08 & 0.90 & \textbf{0.70} & 0.02 & 0.97 \\
         & 41--50 & 0.53 & 0.05 & 0.91 & \textbf{0.69} & 0.02 & 0.95 \\
        \bottomrule
    \end{tabular}
\end{table*}
 
\textbf{Post-processing amplification.} The EWS converts continuous probabilities into three tiers using percentile thresholds ($\approx$0.80 and $\approx$0.39), meaning students with vastly different success probabilities receive identical interventions. The Medium Risk category exhibits poor calibration (Brier Score: 0.18 vs.\ 0.04 for Low Risk); approximately 21\% of High Risk students ultimately succeed. These thresholds amplify demographic disparities beyond those in raw predictions. Unsuccessful male students are 10 percentage points more likely to be categorized High Risk than females (74\% vs.\ 63\%, $p=1.56\times 10^{-13}$), corresponding to roughly 300 additional male students per year receiving intensive interventions. Students $\leq$25 have a much higher probability of High Risk categorization than those 36+ (94\% vs.\ 75\%), and unsuccessful international students are 1.12$\times$ more likely to receive High Risk classification than domestic students, exceeding the 1.08$\times$ difference in raw model accuracy. The EWS additionally operationalizes ``risk'' as probability of non-completion, conflating dropout, transfer, and program change; percentile categorization further weakens construct validity by treating prediction uncertainty (Medium Risk) as equivalent to moderate intervention need, institutionalizing model error as a resource-allocation rule. This amplification is the technical site at which the ASP-HEI cycle's formalization claim becomes observable: the gap between groups is not a property of the model's predictions but of the decision to translate those predictions into a fixed-quota intervention policy.
 
\section{Discussion}\label{discussion}
% Our audit substantiates the central claim of ASP-HEI \S5.4.2: the deployed EWS formalizes existing demographic inequities into institutional policy. More than that, each stage of the ML pipeline is a distinct technical site at which an element of the cycle operates, and at each site inequity is measurably produced rather than merely reflected. The implication is that the pipeline is the cycle instantiated in code.
Each stage of the ML pipeline is a distinct technical site at which an element of the ASP-HEI cycle operates, and at each site inequity is measurably produced rather than merely reflected: the pipeline is the cycle instantiated in code.
 
\textbf{Stages 1--2 --- Historical inequity becomes formal decision rule.} The cycle predicts that algorithmic decision-making learns from a history shaped by pre-existing inequity and converts that diffuse condition into a coded claim \cite{mcconvey_this_2024}. Baseline gaps of 14--18 percentage points in completion between female and male domestic students and between international and domestic populations are not errors in the data; they are the college's institutional history, and an EWS that optimizes for accuracy is asked to preserve it. The conversion to formal claim happens concretely in the model: false-positive rates for domestic students aged 36--40 exceed 0.60, and the model encodes age as a proxy for risk without any causal account of why. Advisors inherit these classifications without access to the underlying reasoning and cannot contest them---exactly the loss of discretionary override the original ethnography attributed to the automation of the faculty--student relationship.
 
\textbf{Stage 3 --- Post-processing as institutional policy.} This is our principal extension of the framework. The cycle argues that algorithmic decisions function as policy once implemented; we identify \emph{percentile-based post-processing} as the specific, locatable mechanism through which this transition happens. Percentile thresholds collapse probabilities from 0.41 to 0.80 into a single ``Medium Risk'' bin---the bin with the worst calibration---and widen the male--female High Risk gap from 36\% to 40\%. This gap is not produced by the model's predictive errors; it is produced by the decision to convert a continuous score into three fixed-quota intervention tiers. A defensible ranking becomes an indefensible allocation rule.
 
\subsection{Three Mechanisms of the Cycle}
 
Locating the mechanism sharpens \emph{why} technical fairness interventions cannot on their own break the cycle. Our findings identify three mechanisms through which the cycle's inequity-exacerbation effect operates, each sharing a common structure: a property of the institution or its tooling is relabeled as a property of the student.
 
\textbf{Task formulation mismatch.} The EWS predicts dropout; advisors interpret its outputs as indicators of who would benefit from support. These are distinct constructs---students drop out for reasons unrelated to academic need (transfer, finances, personal circumstances), and students who persist may still struggle. The mismatch is not a technical oversight; ``who needs help'' has quietly been replaced by ``who threatens retention metrics.''
 
\textbf{Institutional priorities versus student-centered success.} The choice of program non-completion as target reflects the cycle's driving force. The EWS was implemented during funding cuts and tuition freezes \cite{mcconvey_this_2024}, and its operational definition of ``success'' follows directly. Students and advisors may understand success as transferable skills, appropriate career paths, or wellbeing---none of which are observable in the institution's student information system, and none of which make it into the model.
 
\textbf{Prediction uncertainty relabeled as moderate need.} The Medium Risk bin is where the model is least confident; practitioners treat it as where students are moderately in need. Model uncertainty is thus laundered into intervention intensity. This is the cycle's characteristic move applied to epistemic rather than demographic content: a hidden institutional property (how much the model does not know) is recoded as a student property (how much support the student requires).
% The shared structure matters: it is why post-hoc fairness patches, operating at the output of the pipeline, cannot escape a cycle that begins at problem formulation. Fairness in this setting is not a property of a single stage but a property of the whole pipeline's relation to the institution that built it.
 
\section{Implications, Limitations, and Conclusion}
% The contribution of this paper is to give the ASP-HEI cycle's inequity claim an empirical grounding. Prior work theorized that algorithmic decision-making in HEIs formalizes existing inequities as part of a broader pattern of institutional power reproduction \cite{mcconvey_this_2024}; we show, with real institutional training data and the deployed system's design specifications, that this effect operates at specific and identifiable stages of the ML pipeline, and we locate percentile-based post-processing as the site at which baseline demographic differences are converted into amplified allocation policy. 
Historical marginalization enters as training-data disparity, is formalized by the model as decision rules, and is institutionalized by post-processing as allocation policy---and at each stage the institution can intervene, if it chooses to.
 
This reframes what trustworthy ML in public higher education needs to look like. Audits should not stop at aggregate statistical metrics; they must trace how disparity is transformed across stages and interrogate whether the predicted target is the construct the institution claims to care about. Concretely, the post-processing amplification we document points to several pipeline-level interventions worth investigating: replacing fixed-quota percentile bins with calibrated probability thresholds, applying group-conditional calibration \cite{dwork_fairness_2012}, or reformulating the task as multi-objective optimization that balances retention prediction against allocation parity. None of these substitute for revisiting the target definition, but each would attenuate the specific amplification mechanism we identify. Our three mechanisms (task formulation mismatch, institutional-priority drift, and uncertainty relabeled as moderate need) generalize beyond the EWS we audit and can serve as a checklist for auditing other institutional risk models.
 
Our analysis is based on a replica trained on the institution's data and documented design protocol, not the production model itself; fidelity is high but not perfect. We lack data on actual intervention delivery or student responses, so the audit evaluates predictions and categorizations rather than downstream effects, and the outcome label conflates dropout, transfer, and program change---a construct-validity concern we foreground rather than a flaw we can correct. Future work should extend this pipeline-level interrogation to the other elements of the ASP-HEI cycle for which we currently have only ethnographic evidence, and pursue longitudinal evaluation of downstream impact on the students the model governs.
 
\section*{Acknowledgments}
We would like to thank our collaborators and study participants at Centennial College for allowing us to conduct this work. Additionally, we thank the anonymous reviewers whose suggestions and comments helped improve this manuscript. Generative AI tools were used as an editing aid during manuscript preparation; all intellectual content, analysis, and conclusions are solely our own.
 
\printbibliography[heading=subbibintoc]
 
\appendix
 
\section{Supplementary Tables and Figures}
 
\begin{figure}[h]
    \centering
    \includegraphics[width=.85\linewidth]{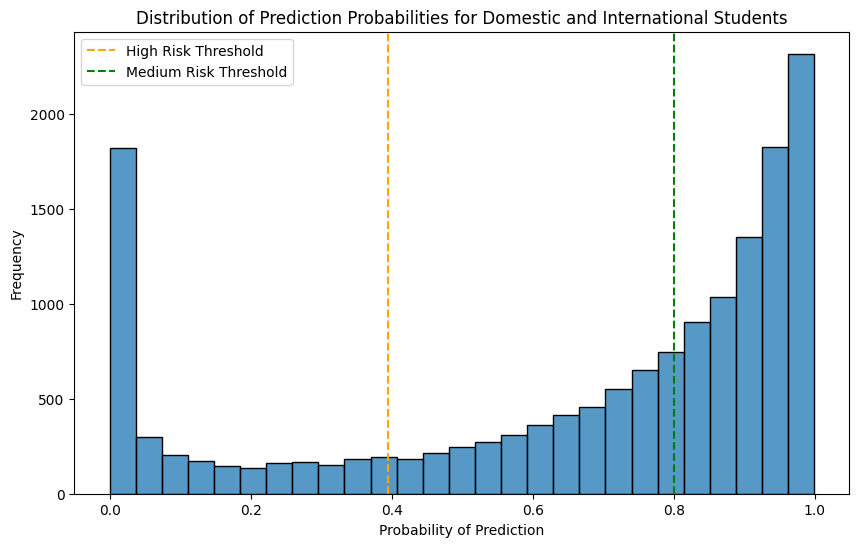}
    \caption{Distribution of success prediction probabilities from the EWS. The orange dashed line ($\approx$0.4) marks the High Risk threshold and the green dashed line ($\approx$0.8) marks the Medium Risk threshold.}
    \label{fig:Distribution of Prediction Probabilities}
\end{figure}

This appendix contains detailed fairness metrics and supplementary visualizations supporting the findings reported in the main text. Figure \ref{fig:Distribution of Prediction Probabilities} shows the bimodal prediction distribution and risk-category thresholds; Tables \ref{tab:fairness_metrics_age_groups} and \ref{tab:international_fairness_metrics_age_groups} give pairwise fairness metrics across age groups for domestic and international students respectively; Table \ref{tab:accuracy_by_risk_level_age_group} reports accuracy by risk level and age group.
 
\begin{table}[h]
\centering
\small
\caption{Domestic fairness metrics across age groups (selected pairs).}
\label{tab:fairness_metrics_age_groups}
\begin{tabular}{lcccc}
\toprule
\textbf{Age Groups} & \textbf{SPD} & \textbf{EOD} & \textbf{AOD} & \textbf{DI} \\
\midrule
21 vs 36 & -0.1710 & -0.1660 & -0.0855 & 0.8034 \\
21 vs 31 & -0.1286 & -0.1447 & -0.0643 & 0.8446 \\
21 vs 41 & -0.1665 & -0.1551 & -0.0833 & 0.8076 \\
21 vs 19 & 0.1669  & 0.1589  & 0.0834  & 1.3137 \\
21 vs 66 & -0.3012 & -0.3971 & -0.1506 & 0.6988 \\
36 vs 19 & 0.3378  & 0.3249  & 0.1689  & 1.6351 \\
31 vs 19 & 0.2954  & 0.3035  & 0.1477  & 1.5554 \\
41 vs 19 & 0.3334  & 0.3140  & 0.1667  & 1.6268 \\
19 vs 66 & -0.4681 & -0.5560 & -0.2340 & 0.5319 \\
\bottomrule
\end{tabular}
\end{table}
 
\begin{table}[h]
\centering
\small
\caption{International fairness metrics across age groups (selected pairs).}
\label{tab:international_fairness_metrics_age_groups}
\begin{tabular}{lcccc}
\toprule
\textbf{Age Groups} & \textbf{SPD} & \textbf{EOD} & \textbf{AOD} & \textbf{DI} \\
\midrule
26 vs 36 & -0.0575 & -0.0402 & -0.0287 & 0.9401 \\
26 vs 19 & 0.1844  & 0.2069  & 0.0922  & 1.2568 \\
21 vs 36 & -0.1050 & -0.0855 & -0.0525 & 0.8906 \\
21 vs 19 & 0.1369  & 0.1616  & 0.0684  & 1.1906 \\
36 vs 19 & 0.2419  & 0.2471  & 0.1209  & 1.3369 \\
31 vs 19 & 0.1771  & 0.1993  & 0.0886  & 1.2467 \\
19 vs 41 & -0.2310 & -0.2108 & -0.1155 & 0.7566 \\
19 vs 51 & -0.2821 & -0.2442 & -0.1410 & 0.7179 \\
\bottomrule
\end{tabular}
\end{table}
 
\begin{table}[h]
    \centering
    \small
    \caption{Accuracy by risk level and age group for domestic and international models, with differences.}
    \label{tab:accuracy_by_risk_level_age_group}
    \begin{tabular}{llccc}
        \toprule
         \textbf{Risk} & \textbf{Age} & \textbf{Dom.} & \textbf{Intl.} & \textbf{Diff.} \\
        \midrule
        \multirow{7}{*}{High}
         & 0  & 0.81 & 0.73 & 0.08 \\
         & 19 & 0.81 & 0.79 & 0.02 \\
         & 21 & 0.84 & 0.72 & 0.12 \\
         & 26 & 0.84 & 0.82 & 0.02 \\
         & 31 & 0.77 & 0.62 & 0.15 \\
         & 36 & 0.65 & 0.60 & 0.05 \\
         & 41 & 0.77 & 0.57 & 0.19 \\
        \midrule
        \multirow{7}{*}{Medium}
         & 0  & 0.70 & 0.83 & -0.14 \\
         & 19 & 0.71 & 0.80 & -0.09 \\
         & 21 & 0.72 & 0.83 & -0.11 \\
         & 26 & 0.74 & 0.79 & -0.05 \\
         & 31 & 0.83 & 0.80 & 0.02 \\
         & 36 & 0.70 & 0.80 & -0.10 \\
         & 41 & 0.77 & 0.73 & 0.04 \\
        \midrule
        \multirow{7}{*}{Low}
         & 0  & 0.91 & 0.97 & -0.06 \\
         & 19 & 0.90 & 0.97 & -0.07 \\
         & 21 & 0.92 & 0.99 & -0.07 \\
         & 26 & 0.94 & 0.98 & -0.04 \\
         & 31 & 0.91 & 0.99 & -0.07 \\
         & 36 & 0.93 & 0.98 & -0.05 \\
         & 41 & 0.92 & 0.95 & -0.03 \\
        \bottomrule
    \end{tabular}
\end{table}
 
\end{document}